\documentclass[aps,twocolumn,groupedaddress,showpacs]{revtex4}
\usepackage[dvips]{graphicx}
\usepackage[]{caption}
\usepackage{amsmath}
\usepackage{amssymb}
\begin{document}
\bibliographystyle{prsty}
\title{Vector Perturbations in a Contracting Universe}
\author{T.~J. Battefeld $^{1)}$ and  R. Brandenberger $^{1,2)}$}
\affiliation{1) Physics Department, Brown University,
  Providence RI 02912 USA.}
\affiliation{2) Perimeter Institute for Theoretical Physics,
35 King Street N., Waterloo, ON N2J 2W9, CANADA.}
\date{\today}
\begin{abstract}
In this note we show that vector perturbations exhibit growing mode solutions in a contracting
Universe, such as the contracting phase of the Pre Big Bang or the Cyclic/Ekpyrotic models of the Universe.
This is not a gauge artifact and will in general lead to the breakdown of perturbation theory -- a severe
problem that has to be addressed in any bouncing model. We also comment
on the possibility of explaining, by means of primordial vector perturbations,
the existence of the observed large scale magnetic fields. This is possible since they can be seeded
by vorticity.
\end{abstract}
\maketitle

\section{Introduction}
Perturbations in the early Universe are a well studied field \cite{bardeen1,MFB,durrer1},
especially in $3+1$ dimensions. It is well known
that vector perturbations (VP) only exhibit decreasing solutions in the context of an expanding Universe,
so most of the literature does not focus on them at all \footnote{See however early work in \cite{Barrow1,Barrow2} in the context of turbulent cosmologies.}.
In recent years a lot of effort has been devoted to studying alternatives to inflationary cosmology,
e.g. the Pre Big Bang scenario (PBB) (for recent reviews see \cite{gasperini1,copeland}),
or the Cyclic/Ekpyrotic model of the Universe \cite{TS,khoury1,tolley1}. These works have focused
on the scalar and tensor metric perturbations. However,
a common feature in these scenarios is that the Universe contracts at early times,
followed by a not well understood bounce. Finally, an expanding Friedmann Universe is reached.
It seems plausible that VP are increasing in a contracting background, so neglecting them totally,
as is usually done in these scenarios, might not be a good idea.

In this note we provide a simple argument as to why VP do indeed grow in a contracting background. We will
show this in a $3+1$ dimensional context where matter is modeled by an ideal fluid. That way our results
will be immediately applicable to the early stages of the PBB scenario, where stringy corrections are
unimportant and an effective $3+1$ dimensional theory is a good approximation. The
application to the Cyclic or Ekpyrotic models of the Universe is more complicated,
since additional dimensions must be taken into account right from the start \cite{tolley1}.
We will also comment on the possibility to seed the observed large scale magnetic fields \cite{GR,W,G}
from vorticities \cite{BDM}.

Our notational conventions are as follows: Latin indices run from 1 to 3, Greek ones from 0 to 3.
Unperturbed quantities carry a $^{(0)}$ on their left and
perturbations have a $\delta$ in front of them. A dot denotes a derivative with respect to
conformal time $\eta$.

\section{Background}
We use a metric with negative signature, scale factor $a(\eta)$, conformal time $\eta$ and,
for simplicity, we consider a flat Universe with metric
\begin{eqnarray}
d\,s^2&=&a^2d\,\eta ^2 -a^2\delta _{ij}d\,x^id\,x^j\,.
\end{eqnarray}
The matter content is modelled by an ideal fluid
\begin{eqnarray}
^{(0)}\!T^\alpha _{\,\,\beta}=(\rho +p)\,^{(0)}\!u^{\alpha}\,^{(0)}\!u_\beta -p\delta ^{\alpha}_{\beta}\,,
\end{eqnarray}
where $p$ is the fluid's pressure, $\rho$ its energy density, and $\,^{(0)}\!u^{\alpha}$ its
four velocity satisfying $\,^{(0)}\!u^{\alpha}\,^{(0)}\!u_\alpha=1$
and given in a comoving frame by $(\,^{(0)}\!u_\alpha)=(a,0,0,0)$. The unperturbed Einstein
equations read
\begin{eqnarray}
^{(0)}\!G^\alpha _{\,\,\beta}=\kappa^2\,^{(0)}\!T^\alpha _{\,\,\beta}\,,
\end{eqnarray}
where $\kappa^2=8\pi G/c^2$, so that
\begin{eqnarray}
2\frac{\ddot{a}}{a}-\left(\frac{\dot{a}}{a}\right)^2&=&-\kappa^2a^2p\,,\\
3\left(\frac{\dot{a}}{a}\right)^2&=&\kappa^2a^2\rho\,,
\end{eqnarray}
yielding for $p=w\rho$ the well known relation $\rho\sim a^{-3(1+w)}$.

\section{Vector perturbations}
We will now write down the most general vector-like metric perturbation of the given background and identify
gauge invariant quantities. After going to Newtonian gauge, we can write down the perturbed
Einstein equations and solve them.

\subsection{Metric}
The most general perturbed metric including only VP is given by \cite{MFB}
\begin{eqnarray}
(\delta g_{\mu\nu})&=&-a^2
\left(\begin{array}{cc}
0&-S^i\\
-S^i&F^i_{\,\,,j}+F^j_{\,\,,i}
\end{array}\right)\,,
\end{eqnarray}
where the vectors $S$ and $F$ are divergenceless, that is $S^{i}_{\,\,,i}=0$ and $F^{i}_{\,\,,i}=0$.
Under a gauge transformation a perturbation $\delta Q$ of a tensor $Q$ transforms as
\begin{eqnarray}
\delta Q \rightarrow {\cal L}_\xi Q
\end{eqnarray}
where ${\cal L}_\xi$ is the Lie derivative. So for
\begin{eqnarray}
x^\mu&\rightarrow& x^\mu +\xi ^\mu\,,\\
(\xi ^\mu)&=&
\left(
\begin{array}{c}
0\\
\xi ^i
\end{array}
\right)\,,
\end{eqnarray}
with $\xi ^{i}_{\,\,,i}=0$ we can compute the change in the perturbed metric tensor, yielding
\begin{eqnarray}
\delta F^i&=&-\xi ^i\,,\\
\delta S^i&=&\dot{\xi}^{i}\,.
\end{eqnarray}
A gauge invariant VP can now be defined as
\begin{eqnarray}
\sigma ^i&=&S^i +\dot{F}^i \,.
\end{eqnarray}
In Newtonian gauge one demands $F^i=0$ so that $\sigma$
coincides with $S$. Note that there is no residual
gauge freedom after going to Newtonian gauge.

\subsection{Energy Momentum Tensor}
The most general perturbation of the energy momentum
tensor including only VP is given by \cite{bardeen1}
\begin{eqnarray}
(\delta T^{\alpha}_{\,\,\beta})&=&
\left(
\begin{array}{cc}
0&-(\rho +p)V^i\\
(\rho +p)(V^i+S^i)&p(\pi ^i_{\,\,,j}+\pi ^j_{\,\,,i})
\end{array}
\right)\,,
\end{eqnarray}
where $\pi ^i$ and $V^i$ are divergenceless. Furthermore $V^i$ is related to the
perturbation in the 4-velocity via
\begin{eqnarray}
(\delta u^\mu)=
\left(
\begin{array}{c}
0\\
\frac{V^i}{a}
\end{array}
\right)\,.\label{defV}
\end{eqnarray}
Gauge invariant quantities are given by
\begin{eqnarray}
\theta ^i&=&V^i-\dot{F}^i
\end{eqnarray}
and $\pi ^i$.

\subsection{Einstein equations}
We will work in Newtonian gauge where $F^i=0$, so our variables are $S^i, V^i$ and $\pi ^i$.
Let us have a look at the perturbed Einstein equations
\begin{eqnarray} \label{pertein}
\delta G^{\alpha}_{\,\,\beta}=\kappa^2\delta T^{\alpha}_{\,\,\beta}\,.
\end{eqnarray}
After some algebra one gets the equations
\begin{eqnarray}
-\frac{1}{2a^2}\bigtriangleup S^i&=&\kappa^2(\rho +p)V^i\,,\label{einstein1}\\
-\frac{1}{2a^4}\partial _t\left(
a^2(S^j_{\,\,,i}+S^i_{\,\,,j})\right)&=&\kappa^2p(\pi ^i_{\,\,,j}+\pi ^j_{\,\,,i})
\,,\label{einstein2}
\end{eqnarray}
where $\bigtriangleup$ is the usual spatial Laplacian
\footnote{These equations
were also derived in \cite{Cartier:2001is} - without, however,
drawing attention to the danger of the growing modes
in the contracting phase}.
With (\ref{einstein1}) we can relate each Fourier mode of the metric perturbation $S_k^i$
to the velocity perturbation $V_k^i$, that is
\begin{eqnarray}
V_k^i=\frac{1}{2\kappa^2a^2(\rho +p)}k^2S_k^i\,.\label{velocitypert}
\end{eqnarray}
Note that only the combination $(\rho +p)V^i$ appears in the energy momentum tensor; therefore
it is this combination that could in principle be observable and may thus be called physically relevant.
 Let us, for simplicity, assume that the right hand side of equation (\ref{einstein2}) vanishes --
this will be the case if there is no anisotropic stress $\pi ^i\equiv0$, or if the Universe is
dominated by pressureless dust $p\equiv 0$.
Then (\ref{einstein2}) yields for any Fourier mode $S_k^i$
\begin{eqnarray}
\partial _t(a^2 S_k^i)=0
\end{eqnarray}
and thus
\begin{eqnarray}
S_k^i=\frac{C_k^i}{a^2}\,, \label{metrikpert}
\end{eqnarray}
where $C_k^i$ is a constant. Combining (\ref{velocitypert}) with (\ref{metrikpert}),
we get for the perturbation of the four velocity
\begin{eqnarray}
V_k^i&=&\frac{k^2}{2\kappa^2(\rho +p)a^2}S_k^i\\
&\sim&\frac{k^2C_k^i}{a^{1-3w}}\label{scaling}\,.
\end{eqnarray}
Thus, in case that the Universe is dominated by dust with $w=0$,
$V_k^i$ is increasing if the scale factor $a$
is decreasing. On the other hand, $V_k^i$ stays constant if the Universe is
dominated by radiation where $w=1/3$.
In contrast, the metric perturbation $S_k^i$ is always increasing, as long as the scale factor decreases.
This feature might be a serious problem in a bouncing Universe as the growth of the vector metric
fluctuations may lead to a breakdown of the validity of the perturbative analysis.

\subsection{Example: Pre Pig Bang}
Let us have a look at the PBB scenario as an example for a bouncing Universe.
In the low curvature PBB regime, the dilaton dominates ($w=1$) and yields
the background solution $a\sim \sqrt{-\eta}$ \cite{gasperini1}, such that,
in the Einstein frame \footnote{Note that the dilaton does not enter into the
Einstein equations (\ref{pertein}) for the vector perturbations.},
\begin{equation}
\sigma _k ^{i} \, = \, S_k^{i}\sim 1/\eta \, .
\end{equation}
At this point, one might start to worry about the validity of the perturbative treatment, an
issue occurring for scalar metric perturbations in this scenario too. In that case it can be
alleviated by working in a gauge in which all perturbations
are at most logarithmically divergent in the limit $k\eta\rightarrow 0$, even though the
gauge invariant variables still diverge (see \cite{gasperini1} or \cite{GGV} for a more
detailed account).

Here the situation is different: Since $\sigma ^i\sim\eta ^{-1}$ and  $\sigma ^i=S^i+\dot{F}^i$ either $S^i$ must
carry the divergent term (which
would render the perturbative treatment invalid at some point) or $\dot{F}^i$. So lets go with the second possibility
such that $F^i\sim \ln(\eta)$ which is tolerable. But if we now recognize that
$\theta ^i\sim a^{-(1-3w)}\sim a^2\sim\eta$ on the one hand and $\theta ^i=V^i-\dot{F}^i$ on the other,
we see that $V^i$ has to cancel out the diverging term $\dot{F}^i\sim\eta ^{-1}$, and thus $V^i\sim\eta ^{-1}$.
So the situation is even worse since now the VP of the energy momentum tensor becomes singular near the bounce.
Thus, in this example, the breakdown of perturbation
theory for vector metric perturbations seems unavoidable.

One might wonder if this puts a new limit on the possible duration of the PBB phase.
This is not the case, since the gauge invariant scalar perturbations already grow
$\sim 1/\eta ^2$ \cite{gasperini1} in this setup.

We have seen that neglecting the VP is by no means justified in a contracting background like in the
PBB scenario. If the scale factor $a$ becomes small enough, backreaktion and non-linear
effects will play an important role. A solution to this problem might be to avoid any matter
content that could exhibit VP, but this seems to be rather artificial.

A more intriguing possibility
to resolve this problem lies in the violation of the null energy condition (NEC) necessary
to get a bounce. At some point $\rho +p$ must become negative and thus $(\rho + p)V^i$
reaches a finite value which implies that $S^i$ has to be finite too. We expect the
violation of the NEC to be of importance only close to the bounce and therefore the question of the validity
of the perturbative analysis remains open \footnote{We thank Cyril Cartier for pointing out this intriguing
possible resolution to the diverging nature of VP near a bounce.}.

Vector perturbations, however, are not only an annoyance one needs to get rid of, but they
might very well help to resolve some open cosmological problems, like the existence of
large-scale magnetic fields, as we shall see in the next section.

\subsection{Possible observable consequences}
Since metric perturbations are not directly observable, we shall focus on the velocity
perturbation $V^i$ defined in (\ref{defV}), and assume that the diverging nature of the
metric perturbation itself at the bounce is resolved by some nonperturbative
mechanism. In order for the velocity perturbation to get excited, we need a matter
source that can exhibit it. The easiest toy model, a scalar field, will obviously not do,
but the next easiest thing, pressureless dust, can have
a velocity perturbation and it might even be observable.

For example, let us assume that the pressureless dust is composed of a two component
cold plasma with different masses, e.g. electrons and protons.
Since $m_p \gg m_e$, we will produce
\begin{eqnarray}
\frac{v^i}{a}:=\frac{1}{2}\left(\delta u^i_{(p)}-\delta u^{i}_{(e)}\right)\neq 0\,.
\end{eqnarray}
Because of (\ref{scaling}), the magnitude of $v^i$ will be increasing until the Universe
becomes radiation-dominated and stays constant thereafter. Let us be optimistic and
assume that this vorticity contrast survives the bounce. Then it should have the same
magnitude during the whole post big bang era until radiation decouples
at a redshift of about $z\sim 1000$. Observations of the CMB constrain the magnitude of
vorticities to be less than
\begin{eqnarray}
\mathcal{K}^i:=\frac{\mbox{curl}\,\left(\frac{v}{a}\right)}{3H}\leq 10^{-5}
\end{eqnarray}
at this redshift \cite{MES}, so we see that the value of $v^i$ in the PBB phase is
constrained by that too.

It has recently been shown that such a small initial vorticity is capable of seeding
large scale magnetic fields \cite{BDM},
which are observed in Galaxies, Clusters, ISM  etc. (see \cite{GR,W,G} for comprehensive reviews).
To be specific, the authors of
\cite{BDM} considered a two-component plasma with an initial $\mathcal{K}^i= 10^{-5}$
at $z=1000$ and computed the magnetic field to be between $10^{-26}$ and $10^{-27}$ G at $z=100$,
large enough for a subsequent dynamo effect to set in.

Here we have given one possible origin for the required initial vorticity in the plasma
\footnote{In \cite{BDM}, the main focus was on an electron-positron plasma, but the results
hold for an electron-proton plasma too, since the corrections are of order unity.}.
One should be aware that the issues of singularity resolution at the bounce (see e.g. \cite{sing}) and the
propagation of the fluctuations through the bounce (see e.g. \cite{bounceflucts}) are
subjects of an ongoing debate,
be it in the context of the PBB scenario or the Cyclic/Ekpyrotic model of the Universe.
Therefore, we will postpone a more detailed analysis.

Nevertheless, we see that the study of VP in a bouncing Universe might yield a novel mechanism
to create magnetic fields \footnote{This is a different
mechanism from what was proposed in \cite{GVG,LL}, where the dilaton seeded magnetic fields.}
without invoking the necessity of breaking the conformal invariance of electromagnetism.
This is only one specific example of how vector perturbations could lead to observable consequences.

One could also look for signals in the CMB -- see \cite{L} for recent work in that direction.

\section{Conclusions}
We have given a simple argument as to why vector perturbations are growing in a contracting Universe.
Near a bounce, it seems unavoidable that the metric perturbations become
non-perturbative in the limit $a\rightarrow 0$, as we saw in the case of the Pre Big Bang scenario.
This issue has to be addressed in any bouncing model of the Universe.
Simply neglecting the vector perturbations is by no means justified.

But besides this annoyance, vector perturbations might have produced observable imprints
to the Universe. One example is the possibility to explain the existence of the large-scale
magnetic fields that have been observed \cite{GR,W,G}, since
these can get seeded by vorticities \cite{BDM}. The details of this mechanism are
model-dependent, e.g. they depend on the nature of the bounce, and we leave
this topic to a future study.

\begin{acknowledgments}
We wish to thank Scott Watson for comments on the manuscript as well as Cyril Cartier and John Barrow
for comments on the first version of this
paper, and also Rocky Kolb for encouraging discussions. TB wishes to acknowledge
the hospitality of the Perimeter Institute. This work was supported
(at Brown Univ.) in part by the
U.S. Department of Energy under Contract DE-FG02-91ER40688, TASK A.
RB thanks the Perimeter Institute for their hospitality and financial
support.
\end{acknowledgments}

\end{document}